\documentclass{cupconf}

\usepackage{epsfig}

  \checkfont{eurm10}
  \iffontfound
    \IfFileExists{upmath.sty}
      {\typeout{^^JFound AMS Euler Roman fonts on the system,
                   using the 'upmath' package.^^J}%
       \usepackage{upmath}}
      {\typeout{^^JFound AMS Euler Roman fonts on the system, but you
                   dont seem to have the}%
       \typeout{'upmath' package installed. cupconf.cls can take advantage
                 of these fonts,^^Jif you use 'upmath' package.^^J}%
      }
  \else
  \fi

  \checkfont{msam10}
  \iffontfound
    \IfFileExists{amssymb.sty}
      {\typeout{^^JFound AMS Symbol fonts on the system, using the
                'amssymb' package.^^J}%
       \usepackage{amssymb}%

      }{}
  \fi

  \IfFileExists{amsbsy.sty}
    {\typeout{^^JFound the 'amsbsy' package on the system, using it.^^J}%
     \usepackage{amsbsy}}
    {\providecommand\boldsymbol[1]{\mbox{\boldmath $##1$}}}

\usepackage{graphicx}

\newcommand\ltsima{$\; \buildrel < \over \sim \;$}
\newcommand\gtsima{$\; \buildrel > \over \sim \;$}
\newcommand\simlt{\lower.5ex\hbox{\ltsima}}
\newcommand\simgt{\lower.5ex\hbox{\gtsima}}

\newsavebox{\astrutbox}
\sbox{\astrutbox}{\rule[-5pt]{0pt}{20pt}}

\newcommand\etal{\mbox{\textit{et al. }}}

\title[Fred Hoyle: Galaxy Formation]{
Fred Hoyle: Contributions to the Theory of Galaxy Formation
  \thanks{Talk presented at the meeting ``New Frontiers in Astronomy''
held at the Institute of Astronomy, Cambridge, on 16th April
2002 in memory
of Fred Hoyle (1915-2001).},\ns}

\author[G. Efstathiou]%
{G. EFSTATHIOU}

\affiliation{Institute of Astronomy, Madingley Road,
Cambridge, CB3 OHA, UK.}

\pubyear{1996}
\volume{538}
\pagerange{1-20}
\date{?? and in revised form ??}
\begin{document}

\maketitle

\begin{abstract}
I review two fundamental contributions that Fred Hoyle made to the
theory of galaxy formation. Hoyle was the first to propose that
protogalaxies acquired their angular momentum via tidal torques from
neighbouring perturbations during a period of gravitational
instability.  To my knowledge, he was also the first to suggest that
the masses of galaxies could be explained by the requirement that
primordial gas clouds cool radiatively on a suitable timescale. Tidal
torques and cooling arguments play a central role in the modern theory
of galaxy formation.  It is a measure of Hoyle's breadth and
inventiveness that he recognised the importance of these processes at
such an early stage in the history of the subject.
\end{abstract}

\firstsection 
\section{Introduction}

I will begin by quoting from an obituary of Sir Fred Hoyle
written by Leon Mestel (2001):

\smallskip

\begin{quote}
`Fred Hoyle was the astrophysicist {\it par excellence}, and much
else. He wrote technical papers on an astonishingly wide range
of astronomical topics, his most important work permanently
widening our vistas and influencing strongly the direction
of future research.'
\end{quote}

\smallskip

\noindent
The `much else' referred to by Leon in this quote includes Hoyle's
contributions to the popularisation of science, prolific science
writing, his creation and directorship of the Institute of Astronomy
at Cambridge and his often visionary work for the U.K. Science Research
Council.

At this meeting our focus has been on Hoyle's astronomical research.
Everyone would agree that Hoyle's outstanding scientific achievements
were in developing the theory of stellar evolution and
in understanding the origin of the chemical elements.  As appropriate,
these topics have taken the central stage at this meeting.  Hoyle's
contributions to galaxy formation, the subject of this article, are
less well known. I have chosen this topic to illustrate the
`astonishingly wide range' of Hoyle's research referred to by Leon Mestel.
I will argue that Hoyle was the first to understand the
masses of galaxies and why galaxies  rotate. These are important results in
their own right and now form an essential part of the modern theory of
galaxy formation. Many astronomers would be proud to list these two
results (which are not even mentioned in any of Hoyle's obituaries) as
their towering achievements!

Section 2 discusses the origin of galactic angular momentum
and is based on Hoyle's (1949) contribution to the quaintly named
volume {\it `Problems of Cosmical Aerodynamics'} edited
by Burgers and van de Hulst. This volume summarises the proceedings
of the equally quaintly named {\it `Symposium on the Motion of 
Gaseous Masses of Cosmical Dimensions'} held in Paris  in 1949.
Section 3 discusses the origin of galactic masses. This is based
on Hoyle's  well known paper {`On the Fragmentation of Gas Clouds
into Galaxies and Stars'} (Hoyle 1953). Curiously, most of the citations to
this paper refer to the sections on star formation
(hierarchical, opacity limited, fragmentation), yet this paper 
provides a clear account of the cooling timescale explanation
of galaxy masses,  predating the work of Binney (1977),
 Rees and Ostriker (1977) and Silk (1977) by more than twenty years.

\section{Origin of Galactic Rotation}\label{sec:rot}

Consider the situation shown in Figure 1. Here we
have an ellipsoid separated by distance $r$ from an object
of mass $M$. Let us assume that the ellipsoid is a homogeneous
oblate spheroid of density $\rho$, mass $M_G$, semi-major axis $a$ and
semi-minor axis $b$. The magnitude of the torque on the spheroid
is ($r \gg a$)
\[
  \Gamma = {3 GM \over 2 r^3}  (\sin 2 \theta) \rho \int (x^2 - z^2) d^3{\bf x}
\]
\begin{equation}
   \qquad \qquad \;\;      = {3 GMQ \over 4 r^3}  \sin 2 \theta, \quad
Q = {2 \over 5} M_G (a^2 - b^2),
  \label{T1}
\end{equation}
where $Q$ is the quadrupole moment of the spheroid.

\begin{figure}
\begin{center}
{\scalebox{0.50}[0.50]{\includegraphics{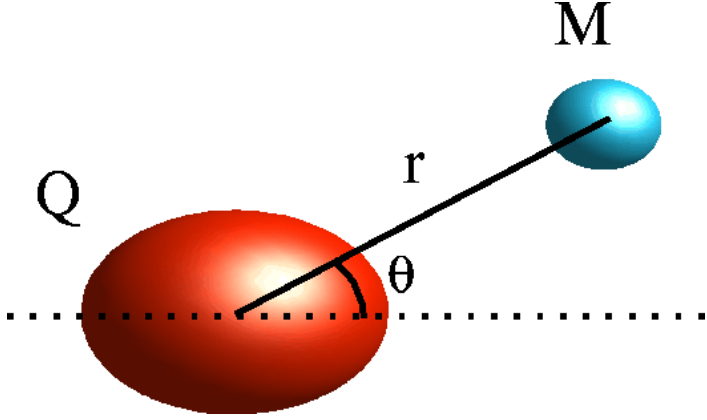}}}
\end{center}
  \caption{An ellipsoid with quadrupole moment $Q$
separated by distance $r$ from a mass $M$.  }\label{fig1}
\end{figure}

How does Hoyle apply equation (\ref{T1})? First, we need to estimate
the timescale over which the torque acts. For this, Hoyle argues that
the torque will act on a timescale comparable to the collapse
timescale of a protogalaxy ($ \sim (3/4 \pi G \rho)^{1/2}$), where
for $\rho$ Hoyle adopts the value $10^{-27} {\rm g/cm}^3$. This was
Hoyle's estimate of the present mean matter density of the Universe
and is high, presumably because estimates of the Hubble constant were
so high at the time (see for example the discussion of the age
discrepancy of evolving world models in Bondi and Gold 1948). We can
re-interpret Hoyle's estimate as the mean density of the Universe at
the redshift at which the protogalaxy begins to form, $(1+z_f) \sim 4
(\Omega_m h^2)^{-1/3}$, which nowadays we would
regard as a reasonable number. Hoyle then assumes that the
protogalaxy collapses until it is centrifugally supported, 
in which case the final angular velocity will be
\begin{equation}
  \Omega_f  \approx \left (  {r^3 \over 3MGx } \right )^3 \left ( 4 \pi G \rho 
\over 3 \right )^{7/2},  \qquad x = {5 Q \over 4 M_G a^2}  \sin 2\theta.
  \label{T2}
\end{equation}
To estimate the term $(r^3/M)$ in (\ref{T2}) Hoyle uses a clever argument.
I quote directly from his paper

\smallskip 

\begin{quote}
`We now reach the important step of interpreting the external
gravitational field that produces the couple acting on the
condensation. Instead of regarding this field as arising from a
neighbouring galaxy, we notice that there are large scale
irregularities in the distribution of the internebular material. The
existence of such irregularities probably exist also among the general
field nebula, as is evidenced by the occurence of peculiar velocities
among these galaxies. Since the peculiar velocities average about $200\;
{\rm km}\;{\rm s}^{-1}$ in the neighborhood of our own galaxy, this
indicates $(GM/r)^{1/2} \approx 2 \times 10^7 \; {\rm cm} \; {\rm s}^{-1}$
in this neighborhood. In addition the observed over-all radii of the great
nebular clusters are of order $10^6$ parsecs, which suggests a value
of $r$ of order $3 \times 10^{24} \;{\rm cm}$.'
\end{quote}

\smallskip

\noindent
Given how little was known about galaxy  clustering at the time, this
is a brilliant way of constraining the matter distribution
and leads to Hoyle's final estimate of
\begin{equation}
  \Omega_f  \approx {10^{-16} \over x^3} \;{\rm s}^{-1},
  \label{T3}
\end{equation}
which he argues is consistent with the characteristic angular velocities
of the Milky Way and Andromeda ($\sim 10^{-15} \; {\rm s}^{-1}$) if the
dimensionless number $x$ is of order $1/3$ (a reasonable value).

What did the conference participants make of this argument? There is a
wonderful discussion reported at the end of Hoyle's contribution,
which I have abstracted:

\smallskip

\begin{itemize}

\item {\it Heisenberg}: I had some difficulty understanding
Dr. Hoyle's argument...  You start with an irregular cloud and then
you ask why this irregular cloud gets an angular momentum. Now, I
don't think that any theory can give us an irregular cloud from the
beginning without giving to it angular momentum at the same
time.... If you have an irregular cloud, then it must have been
produced by some kind of astronomical turbulence.\footnote{By $1949$
Heisenberg's research interests had shifted to the study of
turbulence. It is rumoured that Heisenberg said that he was looking
forward to discussing quantum mechanics with other physicists in
Heaven after he was dead. However, he thought that he would need a
personal audience with God to understand turbulence.}

\item {\it Hoyle}: I take it that Professor Heisenberg assumes turbulence
to be present in the intergalactic medium arising from some unknown source.
For my part I would regard the gravitational forces as the basic 
phenomena. The energy released by the gravitational contraction might 
possibly furnish turbulent motions and might lead to the appearance of
eddies. I feel very doubtful, however, about the assumption that the
intergalactic medium should already have turbulence from itself.

\item {\it Heisenberg}: May I express my view in the following way. A cloud means turbulence and thus you should not start with a sphere without turbulence, since anything which is a cloud is turbulent by itself.

\item {\it Batchelor}: [Who understood Kelvin's circulation theorem] Why?

\item {\it Heisenberg}: How can an irregular thing like a cloud have originated
other than as a consequence of turbulent motion?

\item {\it Hoyle}: A cloud can form in a more or less uniform medium
through gravitational instability.

\end{itemize}

\smallskip

\begin{figure}
\begin{center}
  $\begin{array}{ll}
    {\mbox{(a)} N_p=58^3, z=0.81} & {\mbox{(b)} N_p=58^3, z=0.75} \\
    \epsfig{file=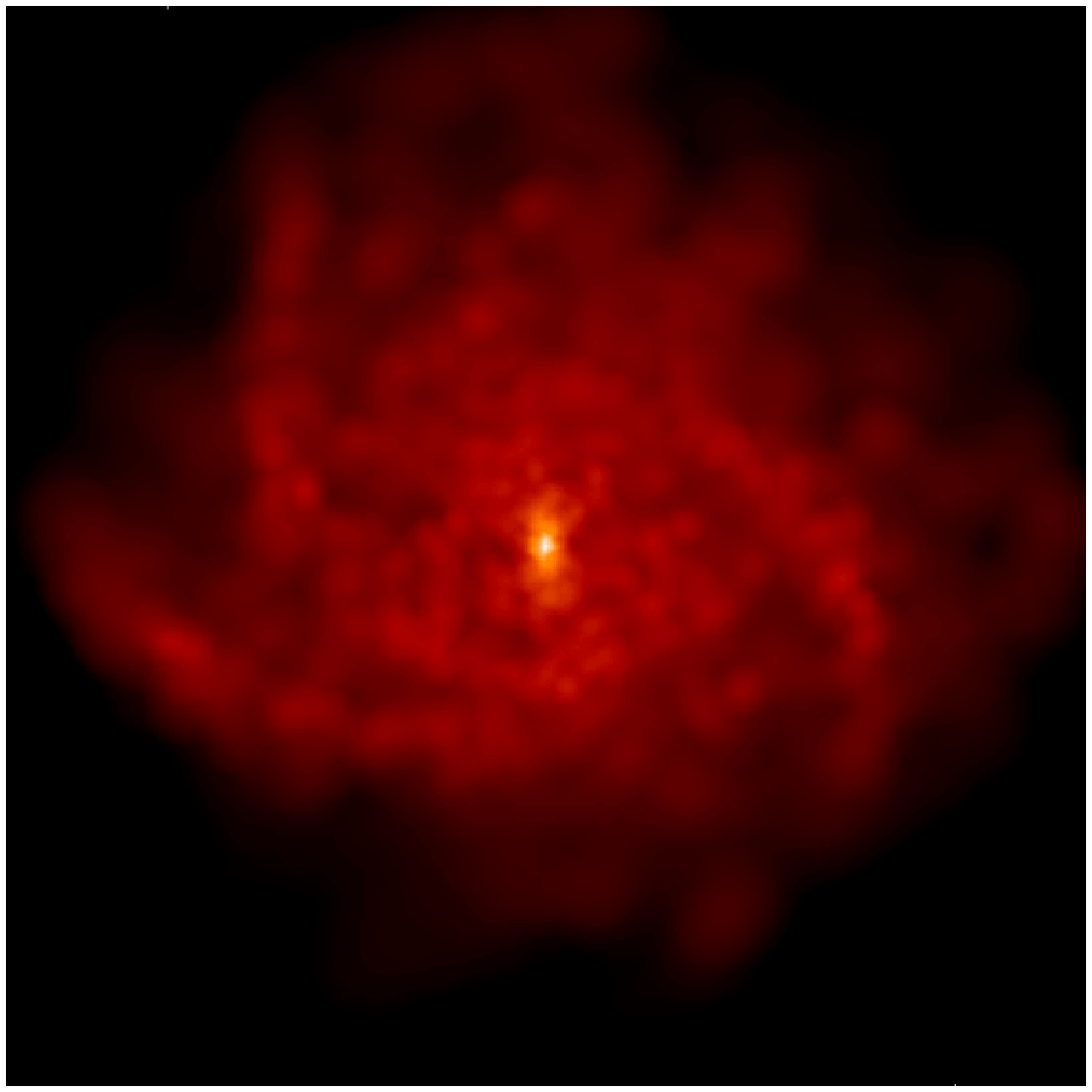,width=60mm} &
    \epsfig{file=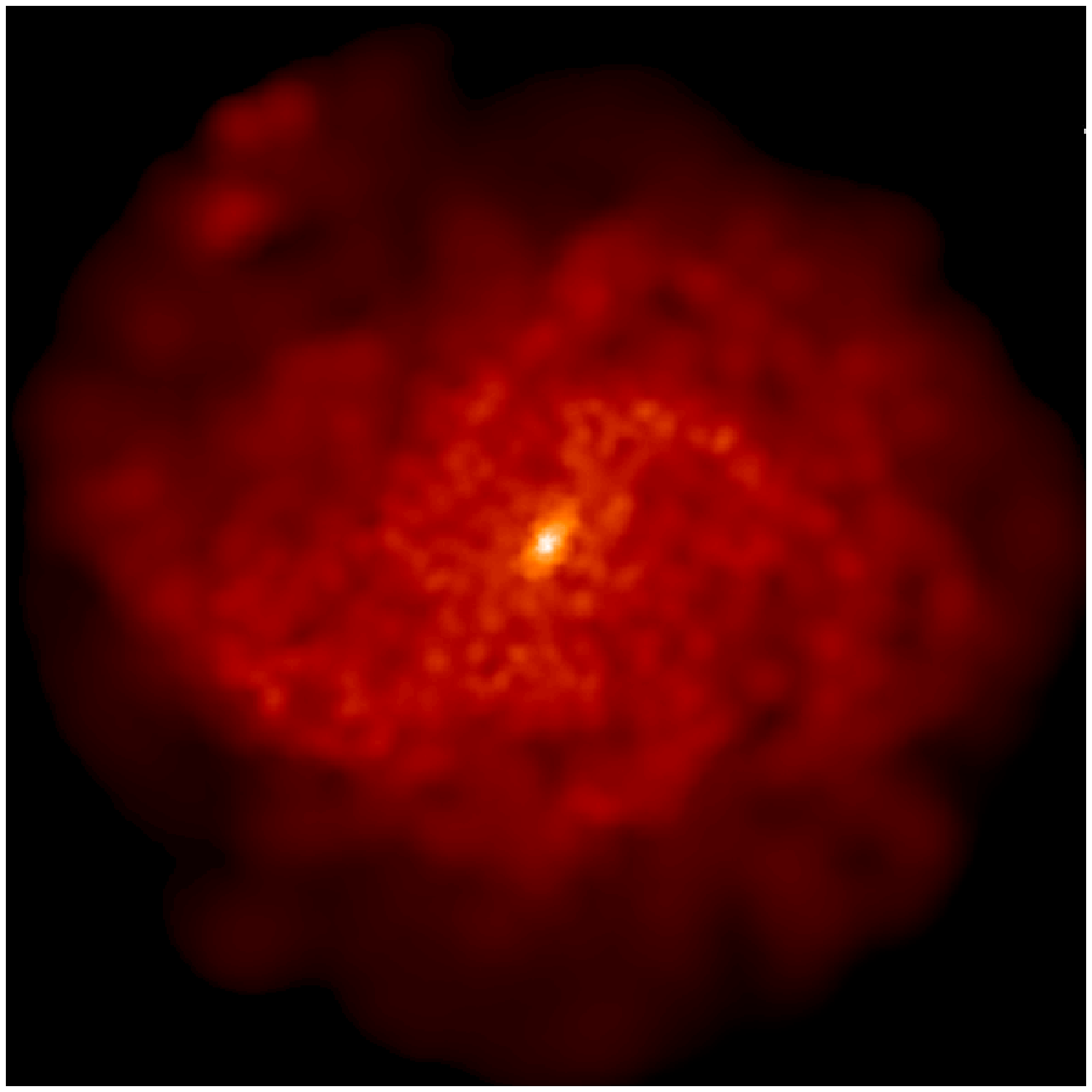,width=60mm}\\
    {\mbox{(c)} N_p=58^3, z=0.70} & {\mbox{(d)} N_p=58^3, z=0.60}  \\
    \epsfig{file=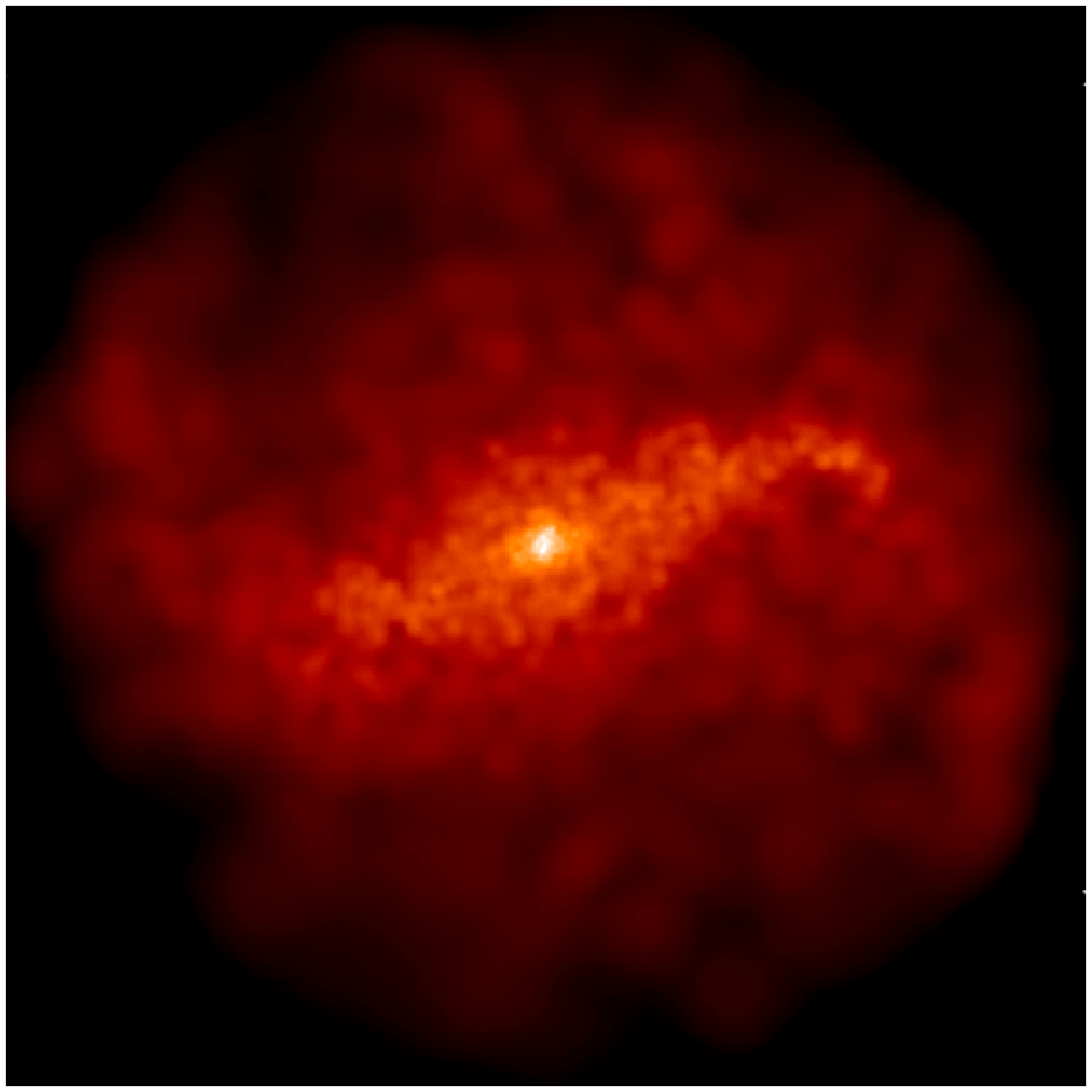,width=60mm}&
    \epsfig{file=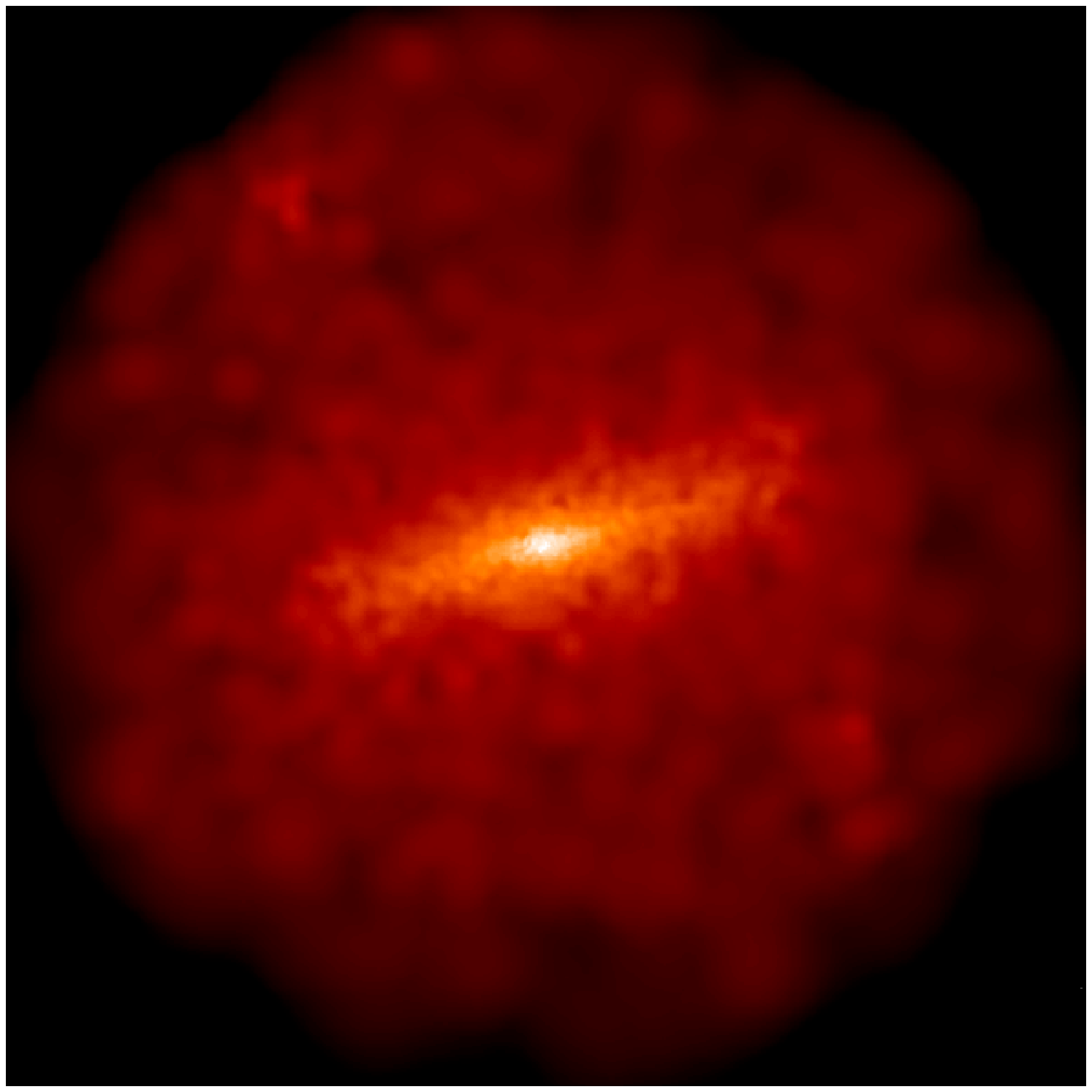,width=60mm}\\
\end{array}$
\caption{Face-on images at four epochs of a disk galaxy formed in 
a gas-dynamical numerical simulation. Each picture shows a square patch
of size $20 \;{\rm kpc} \times 20\; {\rm kpc}$. \label{fig2}}
\end{center}
\end{figure}

I can find nothing of signficance on the tidal torque theory in the
literature (apart from Sciama's, 1955, application to galaxy formation in the
Steady State theory) until Peebles' important paper in 1969. In fact,
the short abstract of Peebles' paper almost echos Hoyle's last remark:

\smallskip
\begin{quote}
`It is shown that the angular momentum of rotation of the Galaxy agrees in magnitude with the prediction of the gravitational instability picture for the
rotation of the galaxies.'
\end{quote}

\smallskip

\noindent
In this paper, Peebles analysed the tidal torque theory using linear
perturbation theory and showed that it could account, roughly, for
the angular momentum of the Milky Way. As part of my thesis work, I used
N-body simulations to estimate the efficiency of the tidal torque mechanism
(Efstathiou and Jones 1979). These simulations suggested that the dimensionless
spin paramater $\lambda$ (roughly the ratio of rotational to kinetic
energy) for a protogalaxy would have a low value of
\begin{equation}
  \lambda = J \vert E \vert^{1/2} G^{-1} M^{-5/2}  \approx 0.05,
  \label{T4}
\end{equation}
a result which was later borne out by much larger numerical
simulations (Barnes and Efstathiou 1987, Zurek, Quinn and Salmon
1988). However, such a low value of $\lambda$ leads to problems with
the theory as envisaged by Hoyle and Peebles, for if a
self-gravitating gas cloud collapses from an initial radius $R_i$ to
form a centrifugally supported exponential disc of scale-length
$\alpha$, it must collapse by a huge factor of 
\begin{equation}
  \alpha R_i  \approx 0.7/ \lambda^2 \approx 300, 
  \label{T5}
\end{equation}
if $\lambda \approx 0.05$. The collapse time for a typical spiral disc
would therefore be longer than the age of the Universe, hence the
theory is untenable. A solution to this problem was proposed by
Efstathiou and Jones (1980) and  worked out in detail by
Fall and Efstathiou (1980). If spiral discs form from the collapse of
baryonic material within a dissipationless dark halo (as envisaged
in the two component theory of White and Rees 1978), and provided
that the gas conserves its angular momentum and is only marginally
self-gravitating when it reaches centrifugal equilibrium, then it
only needs to collapse by a factor of 
\begin{equation}
  \alpha R_i  \approx \sqrt 2/ \lambda \approx 30, 
  \label{T6}
\end{equation}
if $\lambda \approx 0.05$.

Does this lead to a viable theory of galaxy formation? Many
cosmologists would probably say yes, but there are some thorny issues
that have not yet been resolved. The first gas dynamics simulations of
galaxy formation showed that much of the gas collapsed into dense
sub-units at early times (as expected from the over-cooling problem
identified by White and Rees 1978 and discussed in more detail in the
next Section). These sub-units lose their orbital angular momentum by
dynamical friction as they merge to form larger sub-units. The end
result is a blob of gas or stars with specific angular momentum one or
two orders of magnitude lower than observed in real disc
galaxies\footnote{This possibility was first pointed out to me by
Peter Goldreich in 1981.} (see, {\it e.g.} Navarro and White 1994,
Navarro and Steinmetz 1997 ). One possible solution to this problem is
to invoke feedback from supernovae to prevent the gas from collapsing
at early times. (There are differing views, for example, Governato {\it et al.}
(2002) argue that the catastrophic angular momentum loss found in
early simulations is caused by their limited numerical resolution.)
Recently, numerical simulations that include simplified models of
stellar feedback have shown that it is possible to make disc systems
with specific angular momenta and scale-lengths comparable to those of
$L^*$ galaxies (Weil, Eke and Efstathiou 1998, Sommer-Larsen, G\"otz
and Portinari 2002, Abadi \etal 2002).  An example is shown in Figure
2 (Wright, Efstathiou and Eke 2003). This simulation begins with
scale-invariant adiabatic initial conditions in a $\Lambda$-dominated
cold dark matter (CDM) cosmology ($\Omega_m = 0.3$, $\Omega_\Lambda =
0.7$). The gas is (artificially) prevented from cooling before a
redshift of unity, but once cooling sets in, the gas collapses to form
a disc within the dark halo preserving a large fraction of its angular
momentum. As one can see from the figure, the disc displays
(transient) spiral arms at early times and eventually forms a strong
bar. This richness of structure is a direct result of the angular
momentum acquired by tidal torques during the early stages of
gravitational instability, just as Hoyle predicted more than fifty
years ago.

\section{Explaining Galaxy Masses}

Shortly after the discovery of the expansion of the Universe, it was
realised that galaxies had a well defined upper luminosity (Hubble and
Humason 1931). The distribution of galaxy luminosities (the galaxy
luminosity function) was subsequently studied by many authors (see
{\it e.g.} Binggeli, Sandage and Tammann 1988) and has recently been
estimated with extremely high precision using the 2dF and SDSS galaxy
redshift surveys (Blanton \etal 2001, Norberg \etal 2002). These
studies show that the galaxy luminosity function is well described by
a Schechter (1976) function and that 90\% of the mean stellar
luminosity density is contained in galaxies spanning the luminosity
range $0.02 \simlt (L/L^*) \simlt 2.5$, where $L^* \approx 2.6 \times
10^{10} L_\odot$ in the $b$-band. Most of the stellar luminosity in
the Universe is therefore confined to galaxies covering a range of
about a hundred or so in luminosity. Furthermore, there is compelling
evidence for a critical stellar mass of about $3\times 10^{10}M_\odot$
(Kauffmann \etal 2002); galaxies of higher mass have roughly similar
central surface brightnesses, independent of mass or luminosity, while
galaxies with lower stellar masses have lower surface brightnesses
with $\langle \mu \rangle \propto M^{0.54}$.

\begin{figure}
\begin{center}
{\rotatebox{-90}{\scalebox{0.50}[0.50]{\includegraphics{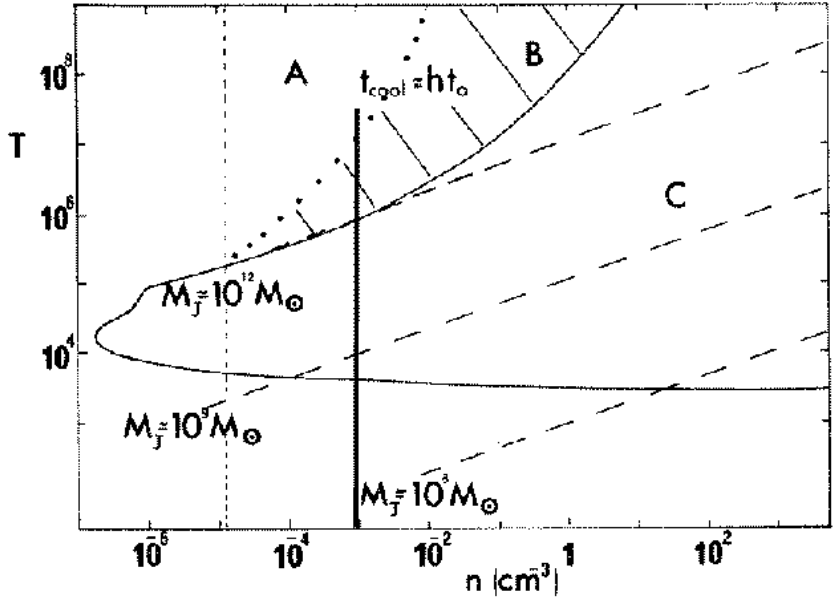}}}}
\end{center}
  \caption{Within region C, bounded by the solid
curve,  a gas cloud of uniform density $n$,  temperature $T$, 
and primordial composition can cool within a free-fall time.
Clouds within  region B can collapse quasi-statically within
a Hubble time. Clouds in region A cannot cool within a Hubble
time. (From Rees and Ostriker,  1977).}
\label{fig3}

\end{figure}

Clearly, some physical process is required to explain the restricted
range of galaxy luminosities and masses. In Hoyle's own words, one of
the aims of his 1953 paper was to explain `Why are the masses of
galaxies mainly confined in the range $3 \times 10^9M_\odot$ to $3
\times 10^{11} M_\odot$, with possibly a tendency to fall into two
groups at the end of this range?'. Hoyle's explanation was based on
simple gas dynamics -- the requirement that a proto-cloud of hydrogen,
at a few times the mean cosmic density, be able to cool radiatively on
a timescale shorter than the timescale for gravitational collapse.

Hoyle's argument can be illustrated in a way more easily accessible to
a modern readers using the famous cooling diagram, reproduced in
Figure 3, from Rees and Ostriker (1977). A gas cloud with a particle
density of $n \sim 10^{-3} {\rm cm}^{-3}$ ($\rho \sim 10^{-27} {\rm
g/cm}^3$, as considered by Hoyle) intersects the solid curve
delineating region C in the diagram at two points, as indicated by the
heavy vertical line in the Figure. Lines of constant Jean's mass (with
slope $1/3$ in this diagram) are shown by the dashed lines. Evidently
a density of $n \sim 10^{-3} {\rm cm}^{-3}$ defines a lower Jean's
mass of $\sim 5 \times 10^{8} M_\odot$ and an upper Jean's mass of
$\sim 10^{12} M_\odot$. The interpretation of this result is as
follows. Region C delineates the region of the $T-n$ plane in which
gas clouds can cool radiatively within a free-fall time. Thus, clouds
with mean density $n \sim 10^{-3} {\rm cm}^{-3}$ heated to their
virial temperatures by gravitational collapse can cool and achieve the
high overdensities ($\gg 10^5$) of normal galaxies only if their
masses lie within the range $ 5 \times 10^{8} M_\odot \simlt M \simlt
10^{12} M_\odot$.  These numbers differ from those in Hoyle's paper
because he neglected cooling by line emission, and he did
not phrase the argument in terms of the virial temperatures of
protoclouds in quite the way discussed above.  The key point, however,
that cooling times delineate upper and lower bounds for galaxy masses,
is contained in Hoyle's paper.

\begin{figure}
\begin{center}
{\rotatebox{0}{\scalebox{0.60}[0.60]{\includegraphics{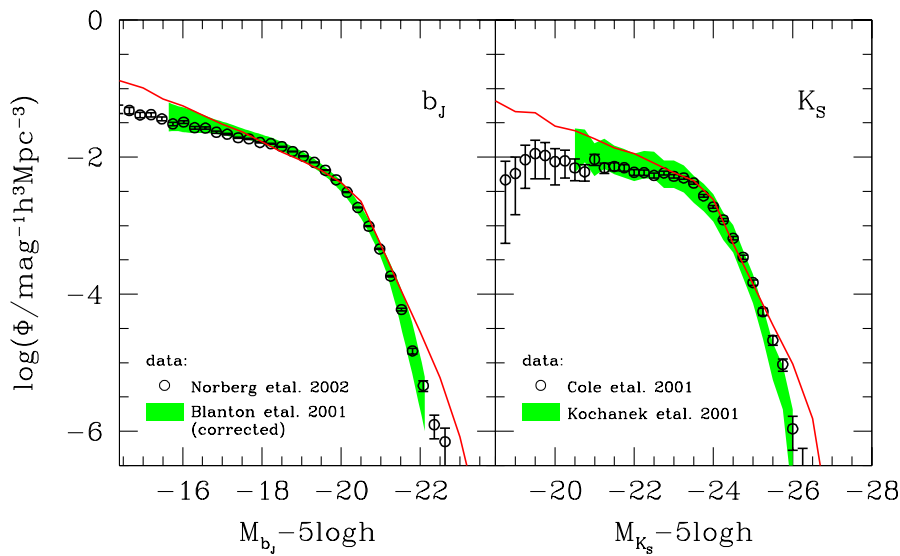}}}}
\end{center}
  \caption{Semi-analytic models of the galaxy luminosity function
from Cole \etal (2000) (solid lines) compared with observations of the
$b$-band  (left-hand panel) and $K$-band (right-hand panel)
luminosity functions. (Figure from Baugh \etal 2002).
  }
\label{fig4}
\end{figure}

Does this cooling argument really explain galaxy masses? Since we no
longer believe in the Steady State theory, we can ask what
happens to clouds with densities much higher than $n \sim 10^{-3}
{\rm cm}^{-3}$. If the virial temperature of such a cloud
exceeds $10^4$K, then it can cool efficiently on a timescale
much shorter than a free-fall timescale. The lower mass limit
of Hoyle's argument is `soft' because it is sensitive to the
redshift at which protogalaxies collapse.

In the cold dark matter model, this leads to a cooling catastrophe, as
mentioned in the previous Section.  In the absence of additional
heating sources, the baryonic material would be expected to collapse
efficiently in small non-linear systems with virial temperatures
$\simgt 10^4$K at high redshifts (White and Rees 1978). According to
the Press-Schechter (1974) theory, in a hierarchical model with a
power-law spectrum of fluctuations $P(k) \propto k^n$, the mass
function of dark haloes at low masses varies as
\begin{equation}
   {dN(m) \over dm} \propto m^{-(9-n)/6}.
  \label{H1}
\end{equation}
For a CDM model, $n \approx -2$ on the scales relevant for galaxy
formation and so the `cooling catastrophe' would lead to a much
steeper mass spectrum than inferred from the faint-end slope of the
galaxy luminosity function ($dN(L)/dL \propto L^{-1.2}$). A
photoionising background can prevent the gas from cooling in haloes
with circular speeds less than about $30 \; {\rm km}\; {\rm s}^{-1}$
(Efstathiou 1992, Benson \etal 2002) and this can help prevent
the formation of dwarf galaxies. However, most authors are agreed
that substantial  feedback from supernovae driven winds is needed to
reproduce the shape of the galaxy luminosity function in CDM models
(see {\it e.g.}  White and Frenk 1991, Cole \etal 2000). As with the
angular momentum problem discussed in the previous section, it seems
that stellar feedback is required to explain the observed
properties of galaxies in a CDM universe.

\begin{figure}
\begin{center}
{\rotatebox{0}{\scalebox{0.70}[0.70]{\includegraphics{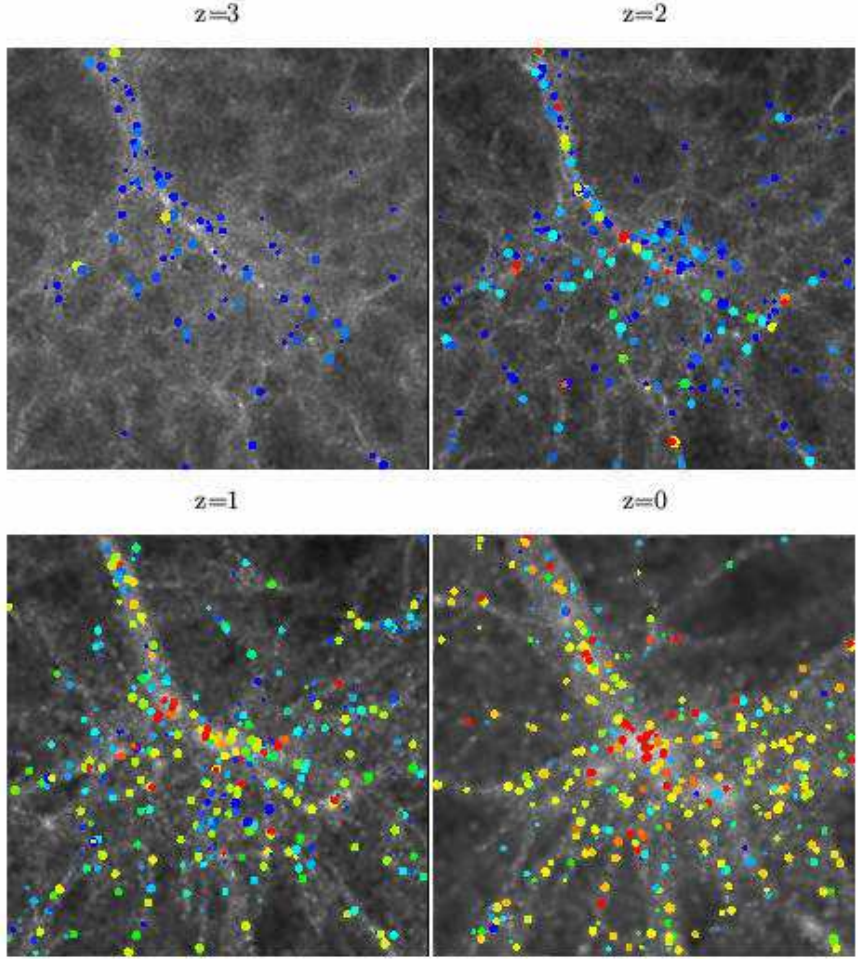}}}}
\end{center}
  \caption{Slices through a $\Lambda$-dominated dark matter simulation
at various redshifts.  Simple prescriptions from semi-analytic models
have been applied to compute the properties of visible galaxies
(masses, stellar ages, {\it etc}) shown by the open circles. (See
Kauffmann \etal 1999).  }\label{fig5}
\end{figure}

This is illustrated in Figure 4 from Baugh \etal (2002). The left hand
panel shows estimates of the $b$-band luminosity function from Blanton
\etal (2001) and Norberg \etal (2002), while the right hand panel
shows estimates of the $K$-band luminosity function from Cole \etal
(2001). The solid lines in the figure show the predictions of the
semi-analytic models of Cole \etal (2000) which include substantial
stellar feedback. These models provide an acceptable match to the
observations, but the model predictions are sensitive to the details
of the feedback model which are poorly know (see {\it e.g.}
Efstathiou, 2000, for a discussion of stellar feedback) and to other
cosmological parameters. For example, the models of Cole \etal (2000)
assume a baryon density of $\Omega_b =0.02$ which is lower than the
value $\Omega_b \approx 0.045$ inferred from recent observations of
the cosmic microwave background radiation (Spergel \etal 2003). A
higher baryon density introduces problems with the bright end of the
luminosity function as well as the faint end (Benson \etal 2003). A
larger baryon density increases the efficiency of radiative cooling leading to
an overproduction of massive galaxies. (This problem is closely related to the
long-standing `cooling flow' problem in cluster cores, see {\it e.g.}
Fabian 2002). The resolution of this problem is still unclear and probably
requires additional sources of feedback, perhaps associated with a
massive central black hole (see {\it e.g.}  Blandford 2001).

In summary, Hoyle correctly identified radiative cooling as an
important process in determining the baryonic masses of
galaxies. However, fifty years after Hoyle's paper, we still do not
understand many of the physical processes that produce galaxies from
an almost featureless spectrum of density fluctuations.  Cooling
clearly plays a role, but so does energy injection from stars and
possibly AGN. The physics of these feedback mechanisms is poorly
understood and yet is crucial for understanding galaxy
formation.  As Figure 5 shows, this physics is vital, for without it
we cannot relate the dark matter distribution to the visible mass
distribution at the present day and at higher redshifts.

\section{Conclusions}

Hoyle's explanation of the rotation of galaxies and his recognition of
the importance of radiative cooling in fixing the baryonic masses of
galaxies are not usually listed amongst his most important
contributions to Astronomy. Yet these processes are central to modern
theories of galaxy formation. It is a measure of Hoyle's breadth and
inventiveness that he recognsed the importance of these processes at
such an early stage in the subject. Nevertheless, more than fifty
years after Hoyle's papers there are still  major gaps in the theory of
galaxy formation. This is because complex physical effects,  in
particular energy injection from massive stars, are important in
determining the baryonic content and angular momentum properties of
galaxies.

I will end with a quotation from Fred Hoyle's book {\it Galaxies, Nuclei
and Quasars}, published in 1966:

\smallskip

\begin{quote}
`It is not too much to say that the understanding of why there are
these different kinds of galaxy, of how galaxies originate, constitutes the
biggest problem in present-day astronomy. The properties of the individual
stars that make up the galaxies form the classical study of astrophysics, while
the phenomena of galaxy formation touches on cosmology. In fact, the study 
of galaxies forms a bridge between conventional astronomy and astrophysics
on the one hand, and cosmology on the other.'
\end{quote}

\smallskip

\noindent
This remains as true today as it was nearly fourty years ago.

\subsection{Acknowledgments}

I think Lisa Wright and Vince Eke for allowing me to reproduce Figure
2 and Carlton Baugh for providing a copy of Figure 4.


\begin{thebibliography}{}

\bibitem[Abadi \etal (2002)]{ANSE02} \textsc{Abadi M.G.,
Navarro J.F., Steinmetz M., Eke V.R.,}  2002, {Simulations of galaxy formation in a 
lambda CDM Universe}, \textit{submitted to ApJ}. (astro-ph/0212282).


\bibitem[Barnes \& Efstathiou  (1987)]{BE87} \textsc{Barnes J.,
Efstathiou G.,} 1987, {Angular momentum from tidal torques}, \textit{ApJ},
\textbf{319}, 575--600.

\bibitem[Baugh \etal (2002)]{B02} 
\textsc{Baugh C.M., Benson A.J.,  Cole S., Frenk C.S.,  Lacey C.}, 2003,
{The evolution of galaxy mass in hierarchical models}
\textit{to appear in `The Mass of Galaxies at Low and High Redshift', 
 eds. R. Bender, A. Renzini}. (astro-ph/0203051).


\bibitem[Benson ]{BLBCF02} \textsc{Benson A.J., Lacey C.G., Baugh
C.M., Cole S., Frenk C.S.,} 2002, {The effects of photoionization on
galaxy formation - I. Model and results at z=0}, \textit{MNRAS},
\textbf{330}, 156--176.


\bibitem[Benson ]{BBFLBC} \textsc{Benson A.J.,  Bower R.G.,
Frenk, C.S.,  Lacey C.G., Baugh C.M.,  Cole S.,}, 2003, 
{What shapes the luminosity function of galaxies?}, \textit{ApJ}
submitted. (astro-ph/0302450).


 \bibitem[Binggeli,  Sandage A., Tammann G. (1988)]{BST88}
     \textsc{Binggeli B., Sandage A., Tammann G.A.,} 1988,
{The luminosity function of galaxies},
\textit{ARAA}, \textbf{26}, 509--560.

 \bibitem[Binney (1977)]{B77}
     \textsc{Binney J.,} 1977,
{The physics of dissipational galaxy formation},
\textit{ApJ}, \textbf{215}, 483--491.


\bibitem[Blandford (2001)]{B01} \textsc{Blandford R.,} 2001, {Energy
release and transport processes in the centres of galaxies}, 
\textit{in `Galaxies and their Constituents at the Highest Angular
Resolutions', Proceedings of IAU Symposium 205, held 15-18 August 2000
at Manchester, United Kingdom, ed R. T. Schilizzi}, 
p10. 


\bibitem[Blanton (2001)]{Betal01}
     \textsc{Blanton M.R., \etal}, 2001,
{The luminosity function of galaxies in SDSS commissioning data},  
\textit{AJ}, \textbf{121}, 2358--2380.

\bibitem[Bondi and Gold (1948)]{BH48}
     \textsc{Bondi H.,  Gold T.,} 1948,
     {The Steady-State theory of the expanding Universe},
     \textit{MNRAS}, \textbf{108}, 252--270.

\bibitem[Cole \etal  (2000)]{C00}
     \textsc{Cole S., Lacey C.G., Baugh C.M., Frenk C.S.,} 2000,
{Hierarchical galaxy formation},
     \textit{MNRAS}, \textbf{319}, 168--204.

\bibitem[Cole \etal  (2001)]{C01}
     \textsc{Cole S., \etal}, 2001,
{The 2dF galaxy redshift survey: near-infrared galaxy luminosity functions},
     \textit{MNRAS}, \textbf{326}, 255--273.

\bibitem[Efstathiou (1992)]{E92}
     \textsc{Efstathiou G.}, 1992,
      {Suppressing the formation of dwarf galaxies via photoionization},
     \textit{MNRAS}, \textbf{256}, 43p--47p.

\bibitem[Efstathiou (2000)]{E00}
     \textsc{Efstathiou G.}, 2000,
      {A model of supernova feedback in galaxy formation},
     \textit{MNRAS}, \textbf{317}, 697--719.

\bibitem[Efstathiou and Jones (1979)]{EJ79} \textsc{Efstathiou G.,
Jones B.J.T.,} 1979, {The rotation of galaxies - numerical investigations
of the tidal torque theory}, \textit{MNRAS}, \textbf{186}, 133--144.

\bibitem[Efstathiou and Jones (1980)]{EJ80} \textsc{Efstathiou G.,
Jones B.J.T.,} 1980, {Angular momentum and the formation of galaxies by
gravitational instability}, \textit{Comments Astrophys. and Space
Science}, \textbf{8}, 169--176.

\bibitem[Fabain (2002)]{F02}
     \textsc{Fabian A.C.}, 2002,
      {Cluster cores and cooling flows},
\textit{to appear in `Galaxy Evolution: Theory and Observations', 
Eds. V. Avila-Reese, C. Firmani, C. Frenk, C. Allen},  in press. (astro-ph/0210150).


\bibitem[Fall and Efstathiou (1980)]{FE80}
     \textsc{Fall S.M., Efstathiou G.}, 1980,
      {Formation and rotation of disc galaxies with haloes},
     \textit{MNRAS}, \textbf{193}, 189--206.

\bibitem[Governato etal (2002)]{Getal02}
     \textsc{Governato F., \etal.}, 2002,
      {The formation of a realistic disk galaxy in lambda dominated cosmologies},
submitted to ApJL. (astro-ph/0207044).

\bibitem[Hoyle (1949)]{H49}
     \textsc{Hoyle F.,} 1949,
{The origin of the rotations of the galaxies,}
\textit{in `Problems of Cosmical Aerodynamics', Proceedings of the Symosium on
the Motion of Gaseous Masses of Cosmical Dimensions held at Paris, August 16-19, 1949. Published by Central Air Documents Office, Ohio,} 195--197.


\bibitem[Hoyle (1953)]{H53}
     \textsc{Hoyle F.,} 1953,
      {On the fragmentation of gas clouds into galaxies and stars},
     \textit{ApJ}, \textbf{118}, 513--528.

\bibitem[Hoyle (1966)]{H66}
     \textsc{Hoyle  F.,} 1966,
      {Galaxies, nuclei and quasars},
     \textit{Heinemann, London.} 



\bibitem[Hubble and Humason (1931)]{HH31} 
\textsc{Hubble E., Humason M.L.,} 1931, 
{The velocity-distance relation among extra-galactic
nebulae}, \textit{ApJ}, \textbf{74}, 43--80.

\bibitem[Kauffmann \etal (1999)]{Ketal99}
     \textsc{Kauffmann G.  Colberg J.M., Antonaldo D., White S.D.M.}, 1999,
      {Clustering of galaxies in a hierarchical universe - I. Methods and results at z=0},
     \textit{MNRAS}, \textbf{303}, 188--206. 

\bibitem[Kauffmann \etal (2002)]{Ketal02}
     \textsc{Kauffmann G. \etal}, 2002,
      {The Dependence of star formation history and internal structure
on stellar mass for $10^5$ low-redshift galaxies},
     \textit{ApJ} submitted. (astro-ph/0205070).

\bibitem[Mestel(2001)]{M01}
     \textsc{Mestel L.,} 2001,
     {Sir Fred Hoyle, FRS, 1915-2001},
     \textit{Astronomy and Geophysics}, \textbf{42}, 5.23-5.24.

\bibitem[Navarro and White (1994)]{NW94}
     \textsc{Navarro J.F., White S.D.M.,} 1994,
     {Simulations of dissipative galaxy formation in hierarchically clustering
universes -- 2. Dynamics of the baryon component in galactic halos},
     \textit{MNRAS}, \textbf{267}, 401-412.

\bibitem[Navarro and Steinmetz (1997)]{NS97}
     \textsc{Navarro J.F., Steinmetz M.,} 1997,
     {The effects of a photoionizing ultraviolet background on the 
formation of disc galaxies},
     \textit{ApJ}, \textbf{478}, 13-28.

\bibitem[Norberg etal (2002)]{Netal02}
     \textsc{Norberg P., \etal}, 2002,
{The 2dF Galaxy Redshift Survey: The $b_J$-band galaxy luminosity function and survey selection function}, \textit{MNRAS}, \textbf{336}, 907--931.

 \bibitem[Peebles(1969)]{P69}
     \textsc{Peebles P.J.E.,} 1969,
{Origin of the angular momentum of galaxies},
     \textit{ApJ}, \textbf{155}, 393--401.

 \bibitem[Press and Schechter (1974)]{PS74} \textsc{Press W.H.,
     Schechter P., } 1974, {Formation of galaxies and clusters of
     galaxies by self-similar gravitational condensation}, \textit{ApJ},
     \textbf{187}, 425--438.

 \bibitem[Rees \& Ostriker (1977)]{RO77}
     \textsc{Rees M.J., Ostriker J.P.,} 1977,
     {Cooling, dynamics and fragmentation of massive gas clouds - Clues to the masses and radii of galaxies and clusters},
     \textit{MNRAS}, \textbf{179}, 541--559.

 \bibitem[Sciama(1955)]{S55}
     \textsc{Sciama D.W.,} 1955,
{On the formation of galaxies in a Steady State universe},
     \textit{MNRAS}, \textbf{115}, 3--14.

 \bibitem[Schechter (1976)]{S76}
     \textsc{Schechter P.}, 1976,
{An analytic expression for the luminosity function for galaxies},
     \textit{ApJ}, \textbf{203}, 297--306.

 \bibitem[Silk(1977)]{S77}
     \textsc{Silk J.,} 1977,
{On the fragmentation of cosmic gas clouds. I - The formation of galaxies and the first generation of stars},
     \textit{ApJ}, \textbf{211}, 638--648.

 \bibitem[Sommer-Larsen, G\"otz and Portinari (2002)]{SLGP02}
     \textsc{Sommer-Larsen J., G\"otz M., Portinari L.,} 2002,
{CDM, feedback and the Hubble sequence},
     \textit{Ap\&SS}, \textbf{281}, 519--524.

 \bibitem[Spergel et al. (2003)]{Setal03} \textsc{Spergel D.N., \etal},
     2003, {First year Wilkinson Microwave Anisotropy Probe (WMAP)
     observations: determination of cosmological parameters},
     \textit{ApJ} submitted. (astro-ph/030220).


 \bibitem[Weil, Eke and Efstathiou (1978)]{WEF98}
     \textsc{Weil M.L., Eke V.R., Efstathiou G.,} 1998,
{The formation of disk galaxies},
     \textit{MNRAS}, \textbf{300}, 773--789.

 \bibitem[White and  Rees (1978)]{WR78}
     \textsc{White S.D.M., Rees M.J.,} 1978,
{Core condensation in heavy halos - 
A two-stage theory for galaxy formation and clustering},
     \textit{MNRAS}, \textbf{183}, 341--358.

 \bibitem[White and  Frenk (1991)]{WF91}
     \textsc{White S.D.M., Frenk C.S.,} 1991,
{Galaxy formation through hierarchical clustering},
     \textit{ApJ}, \textbf{379}, 52--79.

 \bibitem[Wright, Efstathiou and Eke (2003)]{WEE03}
     \textsc{Wright L., Efstathiou G., Eke V.R.,} 2003,
\textit{in preparation}.




\bibitem[Zurek Quinn and Salmon(1988)]{SQS88} \textsc{Zurek W.H.,
Quinn P.J., Salmon J.K.}, 1988, {Rotation of halos in open and closed
universes - differentiated merging and natural selection of galaxies},
\textit{ApJ}, \textbf{330}, 519--534.



\end{thebibliography}
\end{document}